\documentclass[proof]{WileyASNA-v1}
\usepackage{amsmath} 
\newcommand{\angstrom}{\text{\normalfont\AA}}

\articletype{Article Type}%

\received{--}
\revised{--}
\accepted{--}

\begin{document}

\title{The JED-SAD model applied to the AGN UV--X-ray correlation}

\author[1]{Barnier S.*}

\author[1]{Petrucci, P.-O.}

\author[1]{Ferreira, J}

\author[2]{Marcel, G}

\authormark{Barnier \textsc{et al}}

\address[1]{\orgdiv{Univ. Grenoble Alpes}, \orgname{CNRS, IPAG}, \orgaddress{\state{38000 Grenoble}, \country{France}}}

\address[2]{\orgdiv{Institute of Astronomy}, \orgname{University of Cambridge}, \orgaddress{\state{Madingley Road, Cambridge CB3 OHA}, \country{United-Kingdom}}}

\corres{*\email{samuel.barnier@univ-grenoble-alpes.fr}}

\abstract{The non linear correlation between the UV and X-ray emission observed in Active Galactic Nuclei remains a puzzling question that challenged accretion models. While the UV emission originates from the cold disk, the X-ray emission is emitted by a hot corona whose physical characteristics and geometry are still highly debated. The Jet Emitting Disk - Standard Accretion Disk (JED-SAD) is a spectral model stemming from self similar accretion-ejection solutions. It is composed of an inner highly magnetized and hot accretion flow launching jets, the JED, and an outer SAD. The model has been successfully applied to X-ray binaries outbursts. The AGN UV X-ray correlation represent another essential test for the JED-SAD model. We use multiple AGN samples to explore the parameter space and identify the regions able to reproduce the observations. In this first paper, we show that JED-SAD model is able to reproduce the UV--X-ray correlation.}


\jnlcitation{\cname{%
\author{Barnier S.}, 
\author{P.-O.. Petrucci}, 
\author{J. Ferreira}, and
\author{G. Marcel}} (\cyear{2023}), 
\ctitle{The JED-SAD model applied to the AGN UV--X-ray correlation}}


\maketitle

\footnotetext{\textbf{Abbreviations:} AGN, Active Galactic Nuclei; JED, Jet Emitting Disk; SAD, Standard Accretion Disk}

\section{Introduction}\label{sec1}

Active Galactic Nuclei (AGN) host super-massive accreting black holes. They are commonly believed to be powered by an optically thick, geometrically thin accretion disk that emits in the optical and UV energy range \citep{shakura1973black}.  Some, however, also present hard X-ray emission that is generally assumed to result from the disk photons Compton scattering off thermal electrons in the so-called hot corona, located somewhere near the black hole. From the spectrum we can extract a few characteristics of this hot corona: electron temperature around or below 100 keV ($kT_e\sim$100 keV) and optical depth of a few ($\tau\sim 1$) \citep{Gierlinski1997,Zdziarski1998,Wilms2006}. A majority of AGN also present a soft X-ray excess, below 2 keV, with respect to the high-energy power-law extrapolation \citep{Walter1993,Gierlinski2004,Bianchi2009}. 

Many questions still remains open. For instance, the nature of the hot corona is still highly debated. Does the hot corona stand above the black hole and illuminates the disk as the lamppost model proposed (e.g. \citealt{Dauser2014}) ? Or is the corona part of the accretion flow itself as an optically thin and hot flow present in the inner region of the disk (e.g. \citealt{Thorne1975}; \citealt{ferreira2006unified,marcel2019unified,Zdziarski2021}) ? \\

The non-linear correlation between the UV and X-ray monochromatic luminosities taken at 2500 $\angstrom$ and 2 keV is of particular interest \citep{Lusso2010,Risaliti2015,Lusso2016,Lusso2020,Zhu2020,Liu2021}. Indeed, this correlation links the emission of the disk and the emission of the hot corona, thus representing an excellent test for any given supermassive black hole accretion model. 

Multiple recent studies have proposed theoretical explanation for the non-linearity of this correlation. For instance, \cite{Kubota2018} developed a broadband spectral model for bright luminous quasars using three different regions, an outer cold disk, a warm comptonizing region, in the middle, producing the soft X-ray excess observed in most AGN, and a hot corona somewhere close to the black hole. The hot corona is fixed in the inner region of the accretion flow. They define thus four different radii, the ISCO, the outer radius of the hot corona and the outer radius of the warm corona, and the outer radius of the accretion flow. They tie the different regions energetically assuming a Novikov-Thorne emissivity. They fit the data from three AGN with different spectral shape akin to a hard, an intermediate and a close to soft X-ray binaries accretion states. Even though these AGN had different mass accretion rate, the luminosity of the hot corona region is always close to 2\% Eddington. By fixing the power in the corona to 2\% Eddington, the geometry (the three radii) can then be recovered from the value of the mass accretion rate and black hole mass. Using this model, they partially cover the UV--X-ray correlation. They however had difficulties to reproduce large UV/X-ray ratios and the highest luminosity objects of the sample from \cite{Lusso2017}. \cite{Arcodia2019} developed a disk model where the corona extends above the standard disk. The disk pressure can be locally dominated by the gas pressure or by the radiation pressure. Assuming a general formula:   $P_{mag} \propto P_{gas}^\mu P_{tot}^{1-\mu}$ with the power-law index $\mu$ taking values between 0 and 1 depending on the alternative viscosity prescription, they can play with different disk-corona energetic couplings. In the more luminous inner regions, where the radiation pressure dominates, the alternative viscosity decreases the power in the corona and the X-ray output decreases. Varying the mass accretion rate increases the size of this radiation pressure dominated region, decreasing the spectral output of colder regions and thus decreasing the slope of the UV--X-ray correlation. They study the impact of the parameters $\mu$ on the slope in the UV--X-ray correlation when the mass accretion rate increase. They found that the slope of the UV--X-ray correlation is best reproduced with a  viscosity prescription $\mu\sim0.5$.
\\

This paper is a first in a series presenting the application of the jet emitting disk standard accretion disk (JED-SAD, \citealt{ferreira2006unified}) model to the UV--X-ray correlation observed in AGN. In section \ref{sec2}, we present the JED-SAD model and how we simulate AGN spectra. In section \ref{sec3}, we present the sample of AGN we gathered as a reference for the comparison. In section \ref{sec4}, we cover the UV--X-ray plane with a grid of spectra of the JED-SAD model.

\section{The JED-SAD model}\label{sec2}

\subsection{Model}

The JED-SAD is a hybrid disk model where the accretion disk is threaded by a large-scale magnetic field. In the regions where the magnetization $\mu(r) = P_{mag}/P_{tot}$ is of the order of unity (with $P_{mag}$ the magnetic pressure and $P_{tot}$ the total pressure, sum of the thermal and radiation pressure), the magnetic hoop stress overcomes both the outflow pressure gradient and the centrifugal forces: self-confined, non-relativistic jets can be produced. In these conditions, the accretion disk is called a Jet Emitting Disk (JED)  and stems from the MHD self-similar solutions of \cite{fer97}. The effect of the jets on the disk structure can be tremendous since the jets torque can efficiently extract the disk angular momentum, significantly increasing the accretion speed. Consequently, for a given accretion rate, a JED has a much lower density in comparison to the standard accretion disk, and is thus able to produce the hard X-ray emission attributed to the hot corona. The parameter space for stationary JED solutions correspond to magnetization $\mu$ in the range [0.1,1]. In the outer region, where the magnetization is small ($\mu\ll1$) a standard accretion disk (SAD, \citealt{shakura1973black}) is present. Such radial stratification of the magnetic field strength has been observed in recent MHD simulations \citep{scepi2020magnetic,jacquemin2021magnetic}, validating our approach. 

\begin{figure*}[t]
\centerline{\includegraphics[width=0.8\textwidth]{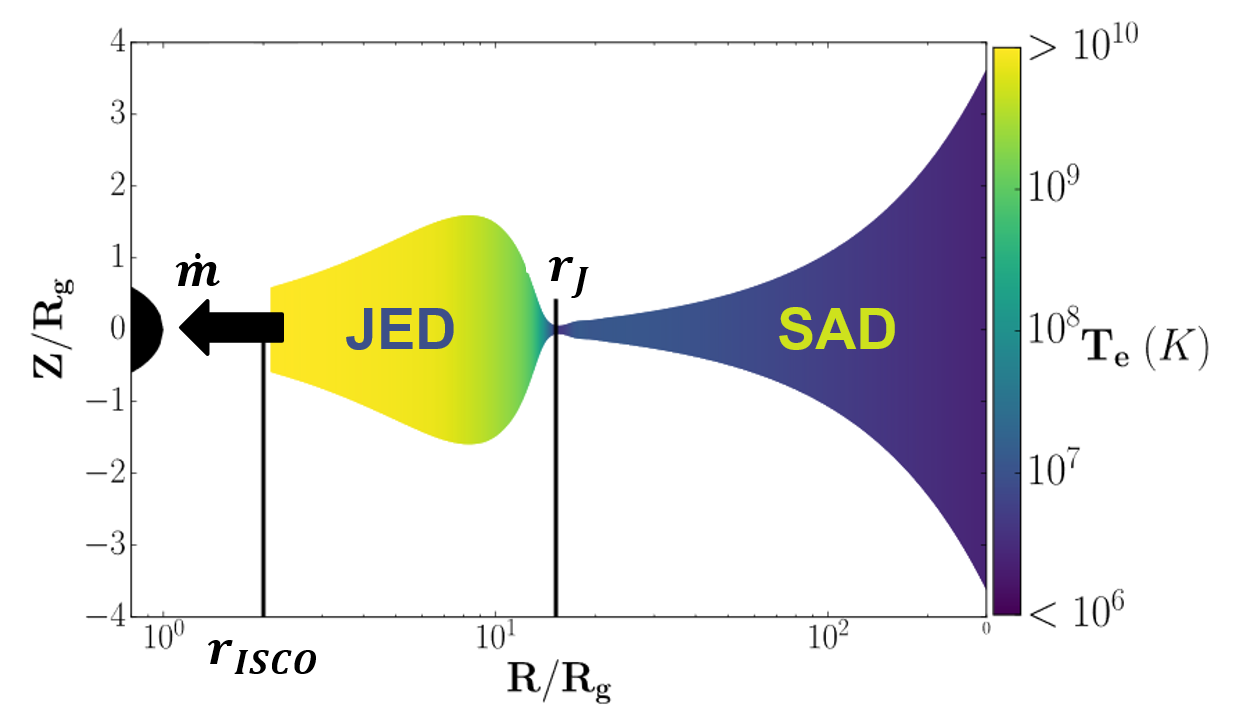}}
\caption{Section of a JED-SAD solution. On the left: the black hole. On the right, the accretion flow. In the inner regions (r<$r_J$), the highly magnetized JED. In the outer regions (r>$r_J$), the SAD. The color represent the mid-plane electronic temperature $T_e$. The JED launches jets, extracting angular momentum, increasing the accretion speed and resulting in a hot and optically thin and geometrically thick accretion flow able to produce hard X-rays.\label{fig1}}
\end{figure*}

\citet{marcel2018bunified} developed a two temperature plasma code solving the thermal equilibrium and computing the spectral energy distribution of any JED-SAD configuration. In Fig. \ref{fig1}, we show an example of the geometry and radial temperature profile of a JED-SAD solution. The complete parameter set of a JED-SAD configuration is defined by:
\begin{itemize}
    \item $\mu$. The mid-plane magnetization inside the JED. Typical values of $\mu$ are within the range [0.1,1].
    \item m. The mass of the black hole in units of solar black hole mass $M_\odot$.
    \item $r_J$. The transition radius between the JED and the SAD region of the accretion flow. $r_J$ is measured in unit of gravitational radius $\displaystyle r_J = R_J / R_G = R_J \cdot c^2 / GM$.
    \item $\dot{m}$. The physical mass accretion rate measured at the inner most stable circular orbit. $\dot{m}$ is in unit of Eddington accretion mass rate $\dot{m}= \dot{M}/\dot{M}_{Edd}= \dot{M} \cdot c^2 / L_{Edd}$. It does not take into account the radiative efficiency and thus $\dot{m}$ can be superior to one without being super-Eddington.
    \item $r_{ISCO}$.  The radius of the inner most stable circular orbit (ISCO). $r_{ISCO}$ is measured in unit of gravitational radius $\displaystyle R_G$. 
    \item $\omega$. A dilution factor depending only on the geometry of the JED-SAD configuration. It controls the number of external photons (produced by the SAD) cooling the JED region and available for comptonisation. $\omega$ is dimensionless and ranges between 0 and 0.5. 
    \item The sonic mach number of the accretion speed, $m_s=u_r/c_s$, with $u_r$ the accretion speed and $c_s$ the local speed of sound. $m_s$ is supposed constant inside the JED and ranges between 0.5 and 3. It is thanks to this supersonic accretion speed that the JED remains optically thin even at high mass accretion rate and can reach the high coronal temperature.
    \item The proportion of the total accretion power inside the JED released in the jets, $b=P_{jets}/P_{acc}$ -- where $P_{jets}$ is the power feeding the jets and $\displaystyle P_{acc}$ the total gravitational power released by the accretion flow within the JED.  $b$ is found between 0.1 and almost 1 for very thin JED \citep{fer97}.
    \item $p$. The index controlling the mass ejected from the JED. The mass accretion rate at a given radius within the jet is then given by $\dot{m}(r)\propto r^{p}$. The JED is characterized by small ejection index $p<0.1$.
\end{itemize}

If the model has already been successfully applied to multiple X-ray binaries outbursts, successfully explaining and reproducing their outbursts using the two main parameters, the mass accretion rate $\dot{m}$ and the transiton radius between the JED and SAD $r_J$ (Cygnus X1 in \citealt{petrucci2010relevance} ; GX339-4 in \citealt{marcel2019unified,Marcel2020,Marcel2022,Barnier22} ; MAXI J1820 in \citealt{marino2021}), applications to AGN are scarce (HE1143 in \citealt{ursini2020nustar}). In the JED-SAD model, the disk and hot corona emissions are computed in a physically motivated configuration, thus, to intensively test the model on AGN, we aim to reproduce the broadband UV--X-ray correlation linking both of these emissions.

\subsection{Spectra}

\begin{figure*}[t]
\centerline{\includegraphics[width=0.8\textwidth]{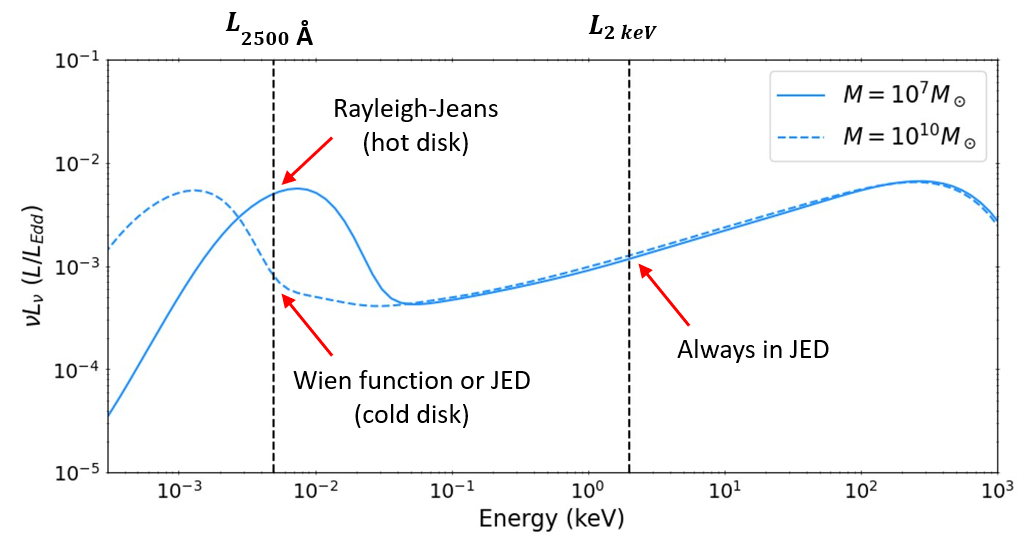}}
\caption{Example of the spectral energy distribution for two different JED-SAD configurations using different black hole masses ($10^7$ in solid line, $10^{10}$ M$_\odot$ in dashed line). To compare the two simulations, we show the luminosity in Eddington units ($L/L_{Edd}$). The bump at low energy is due to the disk blackbody from SAD. In the X-rays, the spectrum is a power-law like emission coming from the JED. The two vertical dashed lines show the energy at which the monochromatic luminosity in the UV (2500\AA) and in the X-ray (2 keV). \label{fig2}}
\end{figure*}

In Fig. \ref{fig2}, we plot the spectra for two JED-SAD solutions using a black hole mass of $10^7$ and $10^{10}$ M$_\odot$. The other JED-SAD parameters are the same for both spectra. We highlight the energies at which the monochromatic luminosities $L_\nu$ are measured for the UV--X-ray correlation. At low energy, the disk blackbody emitted by the SAD peaks in the UV. One can show that the peak temperature of a SAD depends on three main parameters: $kT_{e, SAD}\propto \dot{m}^{1/4}\;r^{-3/4}\;m^{-1/4}$. When the disk is hot (see the $10^7$ M$_\odot$ line for instance in Fig. \ref{fig2}), the UV luminosity is measured within the sum of the Rayleigh-Jeans profile of the different temperature blackbody emitted by the standard disk. When the disk is cold (see the $10^{10}$ M$_\odot$ line in Fig. \ref{fig2}), the UV luminosity can be measured either within the Wien function of the blackbody or even within the JED emission in very cold disk. As such, except as long as the disk is not very cold, the monochromatic luminosity is analytical. The 2 keV monochromatic luminosity is however always measured within the optically thin JED and one must thus solve the thermal equilibrium to get the X-ray monochromatic luminosity.

\section{Samples}\label{sec3}

We selected three samples as reference to compare to our simulations. 

\begin{figure*}[t]
    \centerline{\includegraphics[width=0.75\textwidth]{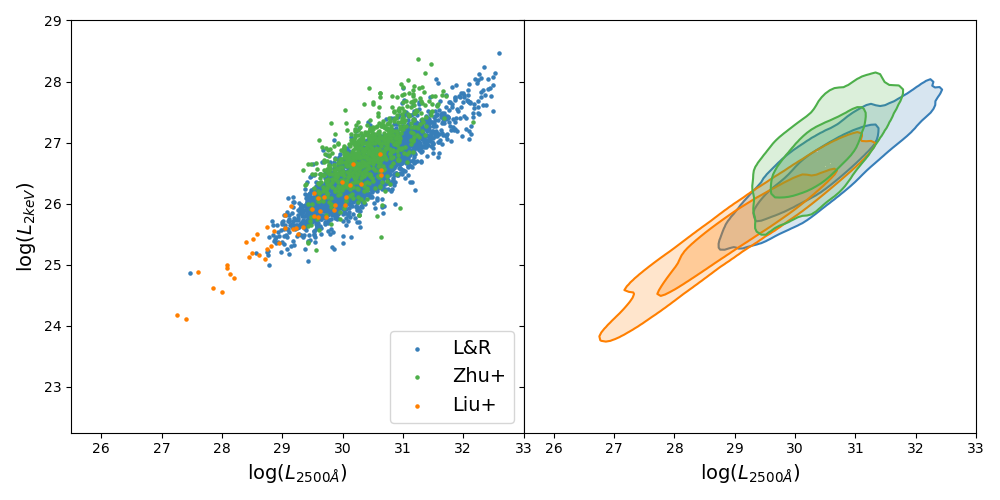}}
    \caption[UV--X-ray correlation: Samples]{The three collected samples. \textbf{Left}: scatter plot of each sample in the UV--X-ray plane. \textbf{Right}: 95\% and 68\% percentile contour for each collected sample in the UV--X-ray plane. Data in blue comes from the work of Lusso \& Risaliti sample \citep{Lusso2020}, in green from the Zhu sample \citep{Zhu2020} and in orange from the Liu sample \citep{Liu2021}. }
    \label{fig3}
\end{figure*}

\paragraph{Lusso \& Risaliti}
The Lusso \& Risaliti sample is the result of an extensive effort to compile an as unbiased as possible AGN sample to study the UV--X-ray correlation with the redshift (e.g. \citealt{Lusso2010,Risaliti2015,Lusso2016,Lusso2017,Risaliti2019,Lusso2020}). They use this correlation to show that quasars can be used as standard candle for cosmology. We will focus on the latest sample \citep{Lusso2020}. The sample consists of 2421 radio-quiet, unabsorbed, bright quasars. They report a correlation $L_{2\; keV}\propto L_{2500\;\angstrom}\,^\gamma$ with index $\gamma=0.586\pm0.061$. In fig. \ref{fig3}, we plot the distribution of the sample in blue. 

One might question the choice of this radio-quiet sample to test the JED-SAD model. Indeed, in the JED-SAD model, we expect the presence of jets which are generally considered as the source of radio emission. However, the nature of the weak radio emission within the radio-quiet population is still highly debated (see review in \citealt{Panessa2019}). Four main scenario are discussed: 1) Radio emission can originate from a strong star formation rate within the host galaxy \citep{Sargent2010, Orienti2015}. 2) The radio emission can come from non-thermal particle accelerated within the magnetised hot corona (e.g. \citealt{Laor2008,Inoue2018,Inoue2021}). 3) Massive winds emitted from the accretion flow can also produce radio emission (see for instance \citealt{Mullaney2013,Zakamska2014,Nims2015,Hwang2018}). 4) Finally, the presence of jetted structure has been observed in a few optically selected local AGN (\citealt{Roy2000,Middelberg2004,Leipski2006}). As such, the JED-SAD model can still be relevant to understand a radio-quiet AGN sample.

\paragraph{Zhu et al.}
In \cite{Zhu2020}, they collected 729 radio loud quasars in order to compare the disk-corona connection obtained for a radio loud population with the radio-quiet case observed in the Lusso \& Risaliti sample. In Fig. \ref{fig3}, we plot the distribution of the sample in green. As reported in \cite{Zhu2020}, the correlation for radio loud quasars is harder compared to the Lusso \& Risaliti sample ($\gamma=0.69\pm0.03$). 

\paragraph{Liu et al.}
\cite{Liu2021} collected 47 AGN with different accretion properties. 21 of them present a super Eddington accretion regime. 26 of them present a sub Eddington regime. They report no difference in correlation between these two populations.  In fig. \ref{fig3}, we plot the distribution of the sample in orange. This sample reaches 2 orders of magnitude lower compared to the Lusso \&Risaliti sample.

\section{Results and prospects}\label{sec4}

\begin{figure}[t]
	\centerline{\includegraphics[width=1\columnwidth]{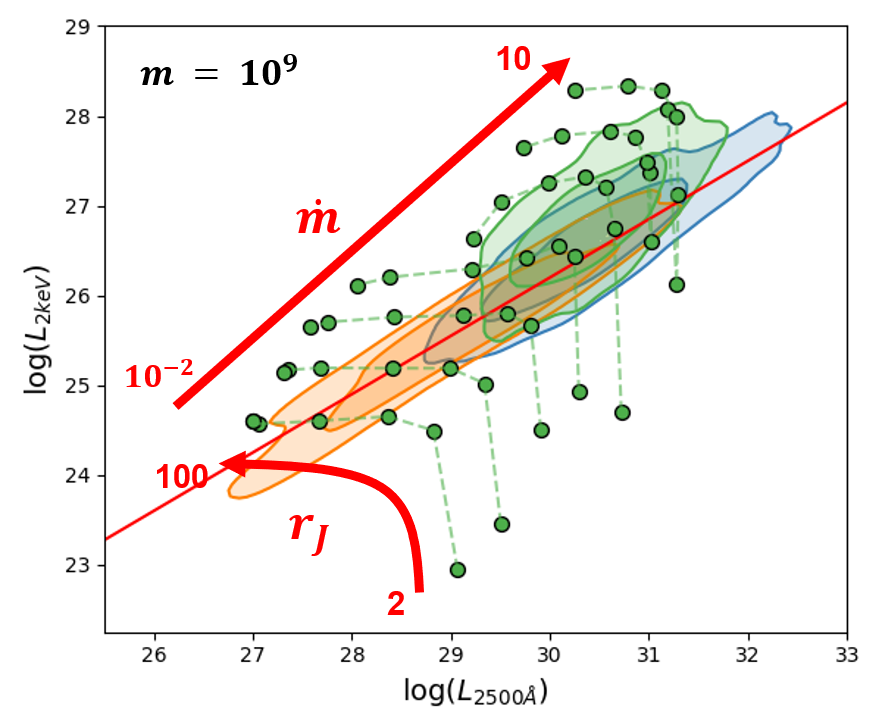}}
	\caption{Grid of JED-SAD spectra in the UV--X-ray plane. Values of $r_J$ go from $2\to100$ (2.1, 4, 8, 15, 27, 52, 100) and of $\dot{m}$ from $10^{-2}\to10$ (0.02, 0.08, 0.2, 0.8, 2, 8). All other parameters are fixed to: $m=10^9$; $r_{isco}=2$; $\omega=0.1$; $m_s=1.5$; $b=0.3$; $p=0.01$. The points connected with a dashed line have the same mass accretion rate. In the background: 95\% and 68\% percentile contour for each sample, using the same color scheme as in Fig. \ref{fig3}.\label{fig4}}
\end{figure}

We produce spectra along a grid of the two main parameters $r_J$ and $\dot{m}$. This grid is build using a logarithmic scale, with $r_J$ varying from 2.1 to 100 and of $\dot{m}$ from $10^{-2}$ to $10$. The other JED-SAD parameters are fixed to: $m=10^9$; $r_{isco}=2$; $\omega=0.1$; $m_s=1.5$; $b=0.3$; $p=0.01$. In Fig. \ref{fig4}, we plot the grid from the simulated AGN JED-SAD spectra over the 95 and 68 percentiles contours of the different collected samples assuming a $10^9$ M$_\odot$ black hole mass. The points connected by a dashed line  use the same value of the mass accretion rate (equi-$\dot{m}$ track). We mark the general evolution with $r_J$ and $\dot{m}$ using red arrows. From the bottom left to the top right, the mass accretion increases, increasing both the UV and X-ray luminosities. From the bottom right to the left of each equi-$\dot{m}$ track, the transition radius increases from the ISCO (2 here) to 100. Indeed, the smaller the transition radius is and the larger the energy released in the standard disk becomes, increasing the UV luminosity.

With this unique grid using a single value of the black hole mass, we see that the JED-SAD simulations are already able to cover almost all of the UV--X-ray samples. Using lower and higher black hole masses, we are able to cover all three complete samples. This show that the JED-SAD parameter space is able to reproduce the UV--X-ray correlation. Furthermore, looking at the straight line of the UV--X-ray correlation in Fig. \ref{fig4}, one can imagine extracting a relation between $r_J$ and $\dot{m}$ that would reproduce the non linearity of the correlation. 

The next step is to study the impact of the other JED-SAD parameters presented in Sec. \ref{sec2}. We can then identify the relevant parameter space of the JED-SAD model allowing to reproduce each of the collected samples and then conclude on the physical origin of the non linear correlation between the UV and X-ray emission in AGN samples. This will be presented in forthcoming papers.

\section*{Acknowledgments}

The authors acknowledge funding support from \fundingAgency{CNES} and the French \fundingAgency{PNHE}.

\bibliography{Wiley-ASNA}%

\begin{thebibliography}{}

\bibitem [\protect \citeauthoryear {%
{Arcodia}%
, {Merloni}%
, {Nandra}%
\BCBL {}\ \BBA {} {Ponti}%
}{%
{Arcodia}%
\ \protect \BOthers {.}}{%
{\protect \APACyear {2019}}%
}]{%
Arcodia2019}
\APACinsertmetastar {%
Arcodia2019}%
\begin{APACrefauthors}%
{Arcodia}, R.%
, {Merloni}, A.%
, {Nandra}, K.%
\BCBL {}\ \BBA {} {Ponti}, G.%
\end{APACrefauthors}%
\unskip\
\newblock
\APACrefYearMonthDay{2019}{{\APACmonth{08}}}{},
\newblock
\unskip
\newblock
\APACjournalVolNumPages{\aap}{628}{}{A135}.
\newblock
\begin{APACrefDOI} \doi{10.1051/0004-6361/201935874} \end{APACrefDOI}
\PrintBackRefs{\CurrentBib}

\bibitem [\protect \citeauthoryear {%
{Barnier}%
\ \protect \BOthers {.}}{%
{Barnier}%
\ \protect \BOthers {.}}{%
{\protect \APACyear {2022}}%
}]{%
Barnier22}
\APACinsertmetastar {%
Barnier22}%
\begin{APACrefauthors}%
{Barnier}, S.%
, {Petrucci}, P\BPBI O.%
, {Ferreira}, J.%
\ et al.\end{APACrefauthors}%
\unskip\
\newblock
\APACrefYearMonthDay{2022}{{\APACmonth{01}}}{},
\newblock
\unskip
\newblock
\APACjournalVolNumPages{\aap}{657}{}{A11}.
\newblock
\begin{APACrefDOI} \doi{10.1051/0004-6361/202141182} \end{APACrefDOI}
\PrintBackRefs{\CurrentBib}

\bibitem [\protect \citeauthoryear {%
{Bianchi}%
, {Guainazzi}%
, {Matt}%
, {Fonseca Bonilla}%
\BCBL {}\ \BBA {} {Ponti}%
}{%
{Bianchi}%
\ \protect \BOthers {.}}{%
{\protect \APACyear {2009}}%
}]{%
Bianchi2009}
\APACinsertmetastar {%
Bianchi2009}%
\begin{APACrefauthors}%
{Bianchi}, S.%
, {Guainazzi}, M.%
, {Matt}, G.%
, {Fonseca Bonilla}, N.%
\BCBL {}\ \BBA {} {Ponti}, G.%
\end{APACrefauthors}%
\unskip\
\newblock
\APACrefYearMonthDay{2009}{{\APACmonth{02}}}{},
\newblock
\unskip
\newblock
\APACjournalVolNumPages{\aap}{495}{2}{421-430}.
\newblock
\begin{APACrefDOI} \doi{10.1051/0004-6361:200810620} \end{APACrefDOI}
\PrintBackRefs{\CurrentBib}

\bibitem [\protect \citeauthoryear {%
{Dauser}%
, {Garcia}%
, {Parker}%
, {Fabian}%
\BCBL {}\ \BBA {} {Wilms}%
}{%
{Dauser}%
\ \protect \BOthers {.}}{%
{\protect \APACyear {2014}}%
}]{%
Dauser2014}
\APACinsertmetastar {%
Dauser2014}%
\begin{APACrefauthors}%
{Dauser}, T.%
, {Garcia}, J.%
, {Parker}, M\BPBI L.%
, {Fabian}, A\BPBI C.%
\BCBL {}\ \BBA {} {Wilms}, J.%
\end{APACrefauthors}%
\unskip\
\newblock
\APACrefYearMonthDay{2014}{{\APACmonth{10}}}{},
\newblock
\unskip
\newblock
\APACjournalVolNumPages{\mnras}{444}{}{L100-L104}.
\newblock
\begin{APACrefDOI} \doi{10.1093/mnrasl/slu125} \end{APACrefDOI}
\PrintBackRefs{\CurrentBib}

\bibitem [\protect \citeauthoryear {%
{Ferreira}%
}{%
{Ferreira}%
}{%
{\protect \APACyear {1997}}%
}]{%
fer97}
\APACinsertmetastar {%
fer97}%
\begin{APACrefauthors}%
{Ferreira}, J.%
\end{APACrefauthors}%
\unskip\
\newblock
\APACrefYearMonthDay{1997}{{\APACmonth{03}}}{},
\newblock
\unskip
\newblock
\APACjournalVolNumPages{Astronomy \& Astrophysics}{319}{}{340--359}.
\PrintBackRefs{\CurrentBib}

\bibitem [\protect \citeauthoryear {%
Ferreira%
, Petrucci%
, Henri%
, Saug{\'e}%
\BCBL {}\ \BBA {} Pelletier%
}{%
Ferreira%
\ \protect \BOthers {.}}{%
{\protect \APACyear {2006}}%
}]{%
ferreira2006unified}
\APACinsertmetastar {%
ferreira2006unified}%
\begin{APACrefauthors}%
Ferreira, J.%
, Petrucci, P\BHBI O.%
, Henri, G.%
, Saug{\'e}, L.%
\BCBL {}\ \BBA {} Pelletier, G.%
\end{APACrefauthors}%
\unskip\
\newblock
\APACrefYearMonthDay{2006}{}{},
\newblock
\unskip
\newblock
\APACjournalVolNumPages{Astronomy \& Astrophysics}{447}{3}{813--825}.
\PrintBackRefs{\CurrentBib}

\bibitem [\protect \citeauthoryear {%
{Gierli{\'n}ski}%
\ \BBA {} {Done}%
}{%
{Gierli{\'n}ski}%
\ \BBA {} {Done}%
}{%
{\protect \APACyear {2004}}%
}]{%
Gierlinski2004}
\APACinsertmetastar {%
Gierlinski2004}%
\begin{APACrefauthors}%
{Gierli{\'n}ski}, M.%
\BCBT {}\ \BBA {} {Done}, C.%
\end{APACrefauthors}%
\unskip\
\newblock
\APACrefYearMonthDay{2004}{{\APACmonth{03}}}{},
\newblock
\unskip
\newblock
\APACjournalVolNumPages{\mnras}{349}{1}{L7-L11}.
\newblock
\begin{APACrefDOI} \doi{10.1111/j.1365-2966.2004.07687.x} \end{APACrefDOI}
\PrintBackRefs{\CurrentBib}

\bibitem [\protect \citeauthoryear {%
{Gierlinski}%
\ \protect \BOthers {.}}{%
{Gierlinski}%
\ \protect \BOthers {.}}{%
{\protect \APACyear {1997}}%
}]{%
Gierlinski1997}
\APACinsertmetastar {%
Gierlinski1997}%
\begin{APACrefauthors}%
{Gierlinski}, M.%
, {Zdziarski}, A\BPBI A.%
, {Done}, C.%
\ et al.\end{APACrefauthors}%
\unskip\
\newblock
\APACrefYearMonthDay{1997}{{\APACmonth{07}}}{},
\newblock
\unskip
\newblock
\APACjournalVolNumPages{\mnras}{288}{4}{958-964}.
\newblock
\begin{APACrefDOI} \doi{10.1093/mnras/288.4.958} \end{APACrefDOI}
\PrintBackRefs{\CurrentBib}

\bibitem [\protect \citeauthoryear {%
{Hwang}%
\ \protect \BOthers {.}}{%
{Hwang}%
\ \protect \BOthers {.}}{%
{\protect \APACyear {2018}}%
}]{%
Hwang2018}
\APACinsertmetastar {%
Hwang2018}%
\begin{APACrefauthors}%
{Hwang}, H\BHBI C.%
, {Zakamska}, N\BPBI L.%
, {Alexandroff}, R\BPBI M.%
, {Hamann}, F.%
, {Greene}, J\BPBI E.%
, {Perrotta}, S.%
\BCBL {}\ \BBA {} {Richards}, G\BPBI T.%
\end{APACrefauthors}%
\unskip\
\newblock
\APACrefYearMonthDay{2018}{{\APACmonth{06}}}{},
\newblock
\unskip
\newblock
\APACjournalVolNumPages{\mnras}{477}{1}{830-844}.
\newblock
\begin{APACrefDOI} \doi{10.1093/mnras/sty742} \end{APACrefDOI}
\PrintBackRefs{\CurrentBib}

\bibitem [\protect \citeauthoryear {%
{Inoue}%
\ \BBA {} {Doi}%
}{%
{Inoue}%
\ \BBA {} {Doi}%
}{%
{\protect \APACyear {2018}}%
}]{%
Inoue2018}
\APACinsertmetastar {%
Inoue2018}%
\begin{APACrefauthors}%
{Inoue}, Y.%
\BCBT {}\ \BBA {} {Doi}, A.%
\end{APACrefauthors}%
\unskip\
\newblock
\APACrefYearMonthDay{2018}{{\APACmonth{12}}}{},
\newblock
\unskip
\newblock
\APACjournalVolNumPages{\apj}{869}{2}{114}.
\newblock
\begin{APACrefDOI} \doi{10.3847/1538-4357/aaeb95} \end{APACrefDOI}
\PrintBackRefs{\CurrentBib}

\bibitem [\protect \citeauthoryear {%
{Inoue}%
, {Khangulyan}%
\BCBL {}\ \BBA {} {Doi}%
}{%
{Inoue}%
\ \protect \BOthers {.}}{%
{\protect \APACyear {2021}}%
}]{%
Inoue2021}
\APACinsertmetastar {%
Inoue2021}%
\begin{APACrefauthors}%
{Inoue}, Y.%
, {Khangulyan}, D.%
\BCBL {}\ \BBA {} {Doi}, A.%
\end{APACrefauthors}%
\unskip\
\newblock
\APACrefYearMonthDay{2021}{{\APACmonth{05}}}{},
\newblock
\unskip
\newblock
\APACjournalVolNumPages{Galaxies}{9}{2}{36}.
\newblock
\begin{APACrefDOI} \doi{10.3390/galaxies9020036} \end{APACrefDOI}
\PrintBackRefs{\CurrentBib}

\bibitem [\protect \citeauthoryear {%
Jacquemin-Ide%
, Lesur%
\BCBL {}\ \BBA {} Ferreira%
}{%
Jacquemin-Ide%
\ \protect \BOthers {.}}{%
{\protect \APACyear {2021}}%
}]{%
jacquemin2021magnetic}
\APACinsertmetastar {%
jacquemin2021magnetic}%
\begin{APACrefauthors}%
Jacquemin-Ide, J.%
, Lesur, G.%
\BCBL {}\ \BBA {} Ferreira, J.%
\end{APACrefauthors}%
\unskip\
\newblock
\APACrefYearMonthDay{2021}{}{},
\newblock
\unskip
\newblock
\APACjournalVolNumPages{Astronomy \& Astrophysics}{647}{}{A192}.
\PrintBackRefs{\CurrentBib}

\bibitem [\protect \citeauthoryear {%
{Kubota}%
\ \BBA {} {Done}%
}{%
{Kubota}%
\ \BBA {} {Done}%
}{%
{\protect \APACyear {2018}}%
}]{%
Kubota2018}
\APACinsertmetastar {%
Kubota2018}%
\begin{APACrefauthors}%
{Kubota}, A.%
\BCBT {}\ \BBA {} {Done}, C.%
\end{APACrefauthors}%
\unskip\
\newblock
\APACrefYearMonthDay{2018}{{\APACmonth{10}}}{},
\newblock
\unskip
\newblock
\APACjournalVolNumPages{\mnras}{480}{1}{1247-1262}.
\newblock
\begin{APACrefDOI} \doi{10.1093/mnras/sty1890} \end{APACrefDOI}
\PrintBackRefs{\CurrentBib}

\bibitem [\protect \citeauthoryear {%
{Laor}%
\ \BBA {} {Behar}%
}{%
{Laor}%
\ \BBA {} {Behar}%
}{%
{\protect \APACyear {2008}}%
}]{%
Laor2008}
\APACinsertmetastar {%
Laor2008}%
\begin{APACrefauthors}%
{Laor}, A.%
\BCBT {}\ \BBA {} {Behar}, E.%
\end{APACrefauthors}%
\unskip\
\newblock
\APACrefYearMonthDay{2008}{{\APACmonth{10}}}{},
\newblock
\unskip
\newblock
\APACjournalVolNumPages{\mnras}{390}{2}{847-862}.
\newblock
\begin{APACrefDOI} \doi{10.1111/j.1365-2966.2008.13806.x} \end{APACrefDOI}
\PrintBackRefs{\CurrentBib}

\bibitem [\protect \citeauthoryear {%
{Leipski}%
, {Falcke}%
, {Bennert}%
\BCBL {}\ \BBA {} {H{\"u}ttemeister}%
}{%
{Leipski}%
\ \protect \BOthers {.}}{%
{\protect \APACyear {2006}}%
}]{%
Leipski2006}
\APACinsertmetastar {%
Leipski2006}%
\begin{APACrefauthors}%
{Leipski}, C.%
, {Falcke}, H.%
, {Bennert}, N.%
\BCBL {}\ \BBA {} {H{\"u}ttemeister}, S.%
\end{APACrefauthors}%
\unskip\
\newblock
\APACrefYearMonthDay{2006}{{\APACmonth{08}}}{},
\newblock
\unskip
\newblock
\APACjournalVolNumPages{\aap}{455}{1}{161-172}.
\newblock
\begin{APACrefDOI} \doi{10.1051/0004-6361:20054311} \end{APACrefDOI}
\PrintBackRefs{\CurrentBib}

\bibitem [\protect \citeauthoryear {%
{Liu}%
\ \protect \BOthers {.}}{%
{Liu}%
\ \protect \BOthers {.}}{%
{\protect \APACyear {2021}}%
}]{%
Liu2021}
\APACinsertmetastar {%
Liu2021}%
\begin{APACrefauthors}%
{Liu}, H.%
, {Luo}, B.%
, {Brandt}, W\BPBI N.%
\ et al.\end{APACrefauthors}%
\unskip\
\newblock
\APACrefYearMonthDay{2021}{{\APACmonth{04}}}{},
\newblock
\unskip
\newblock
\APACjournalVolNumPages{APJ}{910}{2}{103}.
\newblock
\begin{APACrefDOI} \doi{10.3847/1538-4357/abe37f} \end{APACrefDOI}
\PrintBackRefs{\CurrentBib}

\bibitem [\protect \citeauthoryear {%
{Lusso}%
\ \protect \BOthers {.}}{%
{Lusso}%
\ \protect \BOthers {.}}{%
{\protect \APACyear {2010}}%
}]{%
Lusso2010}
\APACinsertmetastar {%
Lusso2010}%
\begin{APACrefauthors}%
{Lusso}, E.%
, {Comastri}, A.%
, {Vignali}, C.%
\ et al.\end{APACrefauthors}%
\unskip\
\newblock
\APACrefYearMonthDay{2010}{{\APACmonth{03}}}{},
\newblock
\unskip
\newblock
\APACjournalVolNumPages{\aap}{512}{}{A34}.
\newblock
\begin{APACrefDOI} \doi{10.1051/0004-6361/200913298} \end{APACrefDOI}
\PrintBackRefs{\CurrentBib}

\bibitem [\protect \citeauthoryear {%
{Lusso}%
\ \BBA {} {Risaliti}%
}{%
{Lusso}%
\ \BBA {} {Risaliti}%
}{%
{\protect \APACyear {2016}}%
}]{%
Lusso2016}
\APACinsertmetastar {%
Lusso2016}%
\begin{APACrefauthors}%
{Lusso}, E.%
\BCBT {}\ \BBA {} {Risaliti}, G.%
\end{APACrefauthors}%
\unskip\
\newblock
\APACrefYearMonthDay{2016}{{\APACmonth{03}}}{},
\newblock
\unskip
\newblock
\APACjournalVolNumPages{APJ}{819}{2}{154}.
\newblock
\begin{APACrefDOI} \doi{10.3847/0004-637X/819/2/154} \end{APACrefDOI}
\PrintBackRefs{\CurrentBib}

\bibitem [\protect \citeauthoryear {%
{Lusso}%
\ \BBA {} {Risaliti}%
}{%
{Lusso}%
\ \BBA {} {Risaliti}%
}{%
{\protect \APACyear {2017}}%
}]{%
Lusso2017}
\APACinsertmetastar {%
Lusso2017}%
\begin{APACrefauthors}%
{Lusso}, E.%
\BCBT {}\ \BBA {} {Risaliti}, G.%
\end{APACrefauthors}%
\unskip\
\newblock
\APACrefYearMonthDay{2017}{{\APACmonth{06}}}{},
\newblock
\unskip
\newblock
\APACjournalVolNumPages{\aap}{602}{}{A79}.
\newblock
\begin{APACrefDOI} \doi{10.1051/0004-6361/201630079} \end{APACrefDOI}
\PrintBackRefs{\CurrentBib}

\bibitem [\protect \citeauthoryear {%
{Lusso}%
\ \protect \BOthers {.}}{%
{Lusso}%
\ \protect \BOthers {.}}{%
{\protect \APACyear {2020}}%
}]{%
Lusso2020}
\APACinsertmetastar {%
Lusso2020}%
\begin{APACrefauthors}%
{Lusso}, E.%
, {Risaliti}, G.%
, {Nardini}, E.%
\ et al.\end{APACrefauthors}%
\unskip\
\newblock
\APACrefYearMonthDay{2020}{{\APACmonth{10}}}{},
\newblock
\unskip
\newblock
\APACjournalVolNumPages{\aap}{642}{}{A150}.
\newblock
\begin{APACrefDOI} \doi{10.1051/0004-6361/202038899} \end{APACrefDOI}
\PrintBackRefs{\CurrentBib}

\bibitem [\protect \citeauthoryear {%
{Marcel}%
\ \protect \BOthers {.}}{%
{Marcel}%
\ \protect \BOthers {.}}{%
{\protect \APACyear {2020}}%
}]{%
Marcel2020}
\APACinsertmetastar {%
Marcel2020}%
\begin{APACrefauthors}%
{Marcel}, G.%
, {Cangemi}, F.%
, {Rodriguez}, J.%
\ et al.\end{APACrefauthors}%
\unskip\
\newblock
\APACrefYearMonthDay{2020}{{\APACmonth{08}}}{},
\newblock
\unskip
\newblock
\APACjournalVolNumPages{\aap}{640}{}{A18}.
\newblock
\begin{APACrefDOI} \doi{10.1051/0004-6361/202037539} \end{APACrefDOI}
\PrintBackRefs{\CurrentBib}

\bibitem [\protect \citeauthoryear {%
{Marcel}%
\ \protect \BOthers {.}}{%
{Marcel}%
\ \protect \BOthers {.}}{%
{\protect \APACyear {2019}}%
}]{%
marcel2019unified}
\APACinsertmetastar {%
marcel2019unified}%
\begin{APACrefauthors}%
{Marcel}, G.%
, Ferreira, J.%
, Clavel, M.%
\ et al.\end{APACrefauthors}%
\unskip\
\newblock
\APACrefYearMonthDay{2019}{}{},
\newblock
\unskip
\newblock
\APACjournalVolNumPages{Astronomy \& Astrophysics}{626}{}{A115}.
\PrintBackRefs{\CurrentBib}

\bibitem [\protect \citeauthoryear {%
{Marcel}%
\ \protect \BOthers {.}}{%
{Marcel}%
\ \protect \BOthers {.}}{%
{\protect \APACyear {2022}}%
}]{%
Marcel2022}
\APACinsertmetastar {%
Marcel2022}%
\begin{APACrefauthors}%
{Marcel}, G.%
, {Ferreira}, J.%
, {Petrucci}, P\BPBI O.%
\ et al.\end{APACrefauthors}%
\unskip\
\newblock
\APACrefYearMonthDay{2022}{{\APACmonth{03}}}{},
\newblock
\unskip
\newblock
\APACjournalVolNumPages{\aap}{659}{}{A194}.
\newblock
\begin{APACrefDOI} \doi{10.1051/0004-6361/202141375} \end{APACrefDOI}
\PrintBackRefs{\CurrentBib}

\bibitem [\protect \citeauthoryear {%
{Marcel}%
\ \protect \BOthers {.}}{%
{Marcel}%
\ \protect \BOthers {.}}{%
{\protect \APACyear {2018}}%
}]{%
marcel2018bunified}
\APACinsertmetastar {%
marcel2018bunified}%
\begin{APACrefauthors}%
{Marcel}, G.%
, Ferreira, J.%
, Petrucci, P\BHBI O.%
\ et al.\end{APACrefauthors}%
\unskip\
\newblock
\APACrefYearMonthDay{2018}{}{},
\newblock
\unskip
\newblock
\APACjournalVolNumPages{Astronomy \& Astrophysics}{617}{}{A46}.
\PrintBackRefs{\CurrentBib}

\bibitem [\protect \citeauthoryear {%
{Marino}%
\ \protect \BOthers {.}}{%
{Marino}%
\ \protect \BOthers {.}}{%
{\protect \APACyear {2021}}%
}]{%
marino2021}
\APACinsertmetastar {%
marino2021}%
\begin{APACrefauthors}%
{Marino}, A.%
, {Barnier}, S.%
, {Petrucci}, P\BPBI O.%
\ et al.\end{APACrefauthors}%
\unskip\
\newblock
\APACrefYearMonthDay{2021}{{\APACmonth{09}}}{},
\newblock
\unskip
\newblock
\APACjournalVolNumPages{accepted A\&A}{}{}{arXiv:2109.00218}.
\PrintBackRefs{\CurrentBib}

\bibitem [\protect \citeauthoryear {%
{Middelberg}%
\ \protect \BOthers {.}}{%
{Middelberg}%
\ \protect \BOthers {.}}{%
{\protect \APACyear {2004}}%
}]{%
Middelberg2004}
\APACinsertmetastar {%
Middelberg2004}%
\begin{APACrefauthors}%
{Middelberg}, E.%
, {Roy}, A\BPBI L.%
, {Nagar}, N\BPBI M.%
\ et al.\end{APACrefauthors}%
\unskip\
\newblock
\APACrefYearMonthDay{2004}{{\APACmonth{04}}}{},
\newblock
\unskip
\newblock
\APACjournalVolNumPages{\aap}{417}{}{925-944}.
\newblock
\begin{APACrefDOI} \doi{10.1051/0004-6361:20040019} \end{APACrefDOI}
\PrintBackRefs{\CurrentBib}

\bibitem [\protect \citeauthoryear {%
{Mullaney}%
\ \protect \BOthers {.}}{%
{Mullaney}%
\ \protect \BOthers {.}}{%
{\protect \APACyear {2013}}%
}]{%
Mullaney2013}
\APACinsertmetastar {%
Mullaney2013}%
\begin{APACrefauthors}%
{Mullaney}, J\BPBI R.%
, {Alexander}, D\BPBI M.%
, {Fine}, S.%
, {Goulding}, A\BPBI D.%
, {Harrison}, C\BPBI M.%
\BCBL {}\ \BBA {} {Hickox}, R\BPBI C.%
\end{APACrefauthors}%
\unskip\
\newblock
\APACrefYearMonthDay{2013}{{\APACmonth{07}}}{},
\newblock
\unskip
\newblock
\APACjournalVolNumPages{\mnras}{433}{1}{622-638}.
\newblock
\begin{APACrefDOI} \doi{10.1093/mnras/stt751} \end{APACrefDOI}
\PrintBackRefs{\CurrentBib}

\bibitem [\protect \citeauthoryear {%
{Nims}%
, {Quataert}%
\BCBL {}\ \BBA {} {Faucher-Gigu{\`e}re}%
}{%
{Nims}%
\ \protect \BOthers {.}}{%
{\protect \APACyear {2015}}%
}]{%
Nims2015}
\APACinsertmetastar {%
Nims2015}%
\begin{APACrefauthors}%
{Nims}, J.%
, {Quataert}, E.%
\BCBL {}\ \BBA {} {Faucher-Gigu{\`e}re}, C\BHBI A.%
\end{APACrefauthors}%
\unskip\
\newblock
\APACrefYearMonthDay{2015}{{\APACmonth{03}}}{},
\newblock
\unskip
\newblock
\APACjournalVolNumPages{\mnras}{447}{4}{3612-3622}.
\newblock
\begin{APACrefDOI} \doi{10.1093/mnras/stu2648} \end{APACrefDOI}
\PrintBackRefs{\CurrentBib}

\bibitem [\protect \citeauthoryear {%
{Orienti}%
, {D'Ammando}%
, {Giroletti}%
, {Giovannini}%
\BCBL {}\ \BBA {} {Panessa}%
}{%
{Orienti}%
\ \protect \BOthers {.}}{%
{\protect \APACyear {2015}}%
}]{%
Orienti2015}
\APACinsertmetastar {%
Orienti2015}%
\begin{APACrefauthors}%
{Orienti}, M.%
, {D'Ammando}, F.%
, {Giroletti}, M.%
, {Giovannini}, G.%
\BCBL {}\ \BBA {} {Panessa}, F.%
\end{APACrefauthors}%
\unskip\
\newblock
\APACrefYearMonthDay{2015}{{\APACmonth{04}}}{},
\newblock
{\BBOQ}\APACrefatitle {{The physics of the radio emission in the quiet side of
  the AGN population with the SKA}} {{The physics of the radio emission in the
  quiet side of the AGN population with the SKA}}.{\BBCQ}
\newblock
\BIn{} \APACrefbtitle {Advancing Astrophysics with the Square Kilometre Array
  (AASKA14)} {Advancing Astrophysics with the Square Kilometre Array
  (AASKA14)}\ \BPG~87.
\PrintBackRefs{\CurrentBib}

\bibitem [\protect \citeauthoryear {%
{Panessa}%
\ \protect \BOthers {.}}{%
{Panessa}%
\ \protect \BOthers {.}}{%
{\protect \APACyear {2019}}%
}]{%
Panessa2019}
\APACinsertmetastar {%
Panessa2019}%
\begin{APACrefauthors}%
{Panessa}, F.%
, {Baldi}, R\BPBI D.%
, {Laor}, A.%
, {Padovani}, P.%
, {Behar}, E.%
\BCBL {}\ \BBA {} {McHardy}, I.%
\end{APACrefauthors}%
\unskip\
\newblock
\APACrefYearMonthDay{2019}{{\APACmonth{04}}}{},
\newblock
\unskip
\newblock
\APACjournalVolNumPages{Nature Astronomy}{3}{}{387-396}.
\newblock
\begin{APACrefDOI} \doi{10.1038/s41550-019-0765-4} \end{APACrefDOI}
\PrintBackRefs{\CurrentBib}

\bibitem [\protect \citeauthoryear {%
{Petrucci}%
, Ferreira%
, Henri%
, Malzac%
\BCBL {}\ \BBA {} Foellmi%
}{%
{Petrucci}%
\ \protect \BOthers {.}}{%
{\protect \APACyear {2010}}%
}]{%
petrucci2010relevance}
\APACinsertmetastar {%
petrucci2010relevance}%
\begin{APACrefauthors}%
{Petrucci}, P\BHBI O.%
, Ferreira, J.%
, Henri, G.%
, Malzac, J.%
\BCBL {}\ \BBA {} Foellmi, C.%
\end{APACrefauthors}%
\unskip\
\newblock
\APACrefYearMonthDay{2010}{}{},
\newblock
\unskip
\newblock
\APACjournalVolNumPages{Astronomy \& Astrophysics}{522}{}{A38}.
\PrintBackRefs{\CurrentBib}

\bibitem [\protect \citeauthoryear {%
{Risaliti}%
\ \BBA {} {Lusso}%
}{%
{Risaliti}%
\ \BBA {} {Lusso}%
}{%
{\protect \APACyear {2015}}%
}]{%
Risaliti2015}
\APACinsertmetastar {%
Risaliti2015}%
\begin{APACrefauthors}%
{Risaliti}, G.%
\BCBT {}\ \BBA {} {Lusso}, E.%
\end{APACrefauthors}%
\unskip\
\newblock
\APACrefYearMonthDay{2015}{{\APACmonth{12}}}{},
\newblock
\unskip
\newblock
\APACjournalVolNumPages{\apj}{815}{1}{33}.
\newblock
\begin{APACrefDOI} \doi{10.1088/0004-637X/815/1/33} \end{APACrefDOI}
\PrintBackRefs{\CurrentBib}

\bibitem [\protect \citeauthoryear {%
{Risaliti}%
\ \BBA {} {Lusso}%
}{%
{Risaliti}%
\ \BBA {} {Lusso}%
}{%
{\protect \APACyear {2019}}%
}]{%
Risaliti2019}
\APACinsertmetastar {%
Risaliti2019}%
\begin{APACrefauthors}%
{Risaliti}, G.%
\BCBT {}\ \BBA {} {Lusso}, E.%
\end{APACrefauthors}%
\unskip\
\newblock
\APACrefYearMonthDay{2019}{{\APACmonth{01}}}{},
\newblock
\unskip
\newblock
\APACjournalVolNumPages{Nature Astronomy}{3}{}{272-277}.
\newblock
\begin{APACrefDOI} \doi{10.1038/s41550-018-0657-z} \end{APACrefDOI}
\PrintBackRefs{\CurrentBib}

\bibitem [\protect \citeauthoryear {%
{Roy}%
, {Wilson}%
, {Ulvestad}%
\BCBL {}\ \BBA {} {Colbert}%
}{%
{Roy}%
\ \protect \BOthers {.}}{%
{\protect \APACyear {2000}}%
}]{%
Roy2000}
\APACinsertmetastar {%
Roy2000}%
\begin{APACrefauthors}%
{Roy}, A\BPBI L.%
, {Wilson}, A\BPBI S.%
, {Ulvestad}, J\BPBI S.%
\BCBL {}\ \BBA {} {Colbert}, J\BPBI M.%
\end{APACrefauthors}%
\unskip\
\newblock
\APACrefYearMonthDay{2000}{{\APACmonth{01}}}{},
\newblock
{\BBOQ}\APACrefatitle {{Slow jets in Seyfert Galaxies: NGC1068}} {{Slow jets in
  Seyfert Galaxies: NGC1068}}.{\BBCQ}
\newblock
\BIn{} J\BPBI E.~{Conway}, A\BPBI G.~{Polatidis}, R\BPBI S.~{Booth}\BCBL {}\
  \BOthers {.}\ (\BEDS), \APACrefbtitle {EVN Symposium 2000, Proceedings of the
  5th european VLBI Network Symposium} {EVN Symposium 2000, Proceedings of the
  5th european VLBI Network Symposium}\ \BPG~7.
\PrintBackRefs{\CurrentBib}

\bibitem [\protect \citeauthoryear {%
{Sargent}%
\ \protect \BOthers {.}}{%
{Sargent}%
\ \protect \BOthers {.}}{%
{\protect \APACyear {2010}}%
}]{%
Sargent2010}
\APACinsertmetastar {%
Sargent2010}%
\begin{APACrefauthors}%
{Sargent}, M\BPBI T.%
, {Schinnerer}, E.%
, {Murphy}, E.%
\ et al.\end{APACrefauthors}%
\unskip\
\newblock
\APACrefYearMonthDay{2010}{{\APACmonth{02}}}{},
\newblock
\unskip
\newblock
\APACjournalVolNumPages{\apjs}{186}{2}{341-377}.
\newblock
\begin{APACrefDOI} \doi{10.1088/0067-0049/186/2/341} \end{APACrefDOI}
\PrintBackRefs{\CurrentBib}

\bibitem [\protect \citeauthoryear {%
Scepi%
, Lesur%
, Dubus%
\BCBL {}\ \BBA {} Jacquemin-Ide%
}{%
Scepi%
\ \protect \BOthers {.}}{%
{\protect \APACyear {2020}}%
}]{%
scepi2020magnetic}
\APACinsertmetastar {%
scepi2020magnetic}%
\begin{APACrefauthors}%
Scepi, N.%
, Lesur, G.%
, Dubus, G.%
\BCBL {}\ \BBA {} Jacquemin-Ide, J.%
\end{APACrefauthors}%
\unskip\
\newblock
\APACrefYearMonthDay{2020}{}{},
\newblock
\unskip
\newblock
\APACjournalVolNumPages{Astronomy \& Astrophysics}{641}{}{A133}.
\PrintBackRefs{\CurrentBib}

\bibitem [\protect \citeauthoryear {%
Shakura%
\ \BBA {} Sunyaev%
}{%
Shakura%
\ \BBA {} Sunyaev%
}{%
{\protect \APACyear {1973}}%
}]{%
shakura1973black}
\APACinsertmetastar {%
shakura1973black}%
\begin{APACrefauthors}%
Shakura, N\BPBI I.%
\BCBT {}\ \BBA {} Sunyaev, R\BPBI A.%
\end{APACrefauthors}%
\unskip\
\newblock
\APACrefYearMonthDay{1973}{}{},
\newblock
\unskip
\newblock
\APACjournalVolNumPages{Astronomy and Astrophysics}{24}{}{337--355}.
\PrintBackRefs{\CurrentBib}

\bibitem [\protect \citeauthoryear {%
{Thorne}%
\ \BBA {} {Price}%
}{%
{Thorne}%
\ \BBA {} {Price}%
}{%
{\protect \APACyear {1975}}%
}]{%
Thorne1975}
\APACinsertmetastar {%
Thorne1975}%
\begin{APACrefauthors}%
{Thorne}, K\BPBI S.%
\BCBT {}\ \BBA {} {Price}, R\BPBI H.%
\end{APACrefauthors}%
\unskip\
\newblock
\APACrefYearMonthDay{1975}{{\APACmonth{02}}}{},
\newblock
\unskip
\newblock
\APACjournalVolNumPages{\apjl}{195}{}{L101-L105}.
\newblock
\begin{APACrefDOI} \doi{10.1086/181720} \end{APACrefDOI}
\PrintBackRefs{\CurrentBib}

\bibitem [\protect \citeauthoryear {%
Ursini%
\ \protect \BOthers {.}}{%
Ursini%
\ \protect \BOthers {.}}{%
{\protect \APACyear {2020}}%
}]{%
ursini2020nustar}
\APACinsertmetastar {%
ursini2020nustar}%
\begin{APACrefauthors}%
Ursini, F.%
, Petrucci, P\BHBI O.%
, Bianchi, S.%
\ et al.\end{APACrefauthors}%
\unskip\
\newblock
\APACrefYearMonthDay{2020}{}{},
\newblock
\unskip
\newblock
\APACjournalVolNumPages{Astronomy \& Astrophysics}{634}{}{A92}.
\PrintBackRefs{\CurrentBib}

\bibitem [\protect \citeauthoryear {%
{Walter}%
\ \BBA {} {Fink}%
}{%
{Walter}%
\ \BBA {} {Fink}%
}{%
{\protect \APACyear {1993}}%
}]{%
Walter1993}
\APACinsertmetastar {%
Walter1993}%
\begin{APACrefauthors}%
{Walter}, R.%
\BCBT {}\ \BBA {} {Fink}, H\BPBI H.%
\end{APACrefauthors}%
\unskip\
\newblock
\APACrefYearMonthDay{1993}{{\APACmonth{07}}}{},
\newblock
\unskip
\newblock
\APACjournalVolNumPages{\aap}{274}{}{105}.
\PrintBackRefs{\CurrentBib}

\bibitem [\protect \citeauthoryear {%
{Wilms}%
, {Nowak}%
, {Pottschmidt}%
, {Pooley}%
\BCBL {}\ \BBA {} {Fritz}%
}{%
{Wilms}%
\ \protect \BOthers {.}}{%
{\protect \APACyear {2006}}%
}]{%
Wilms2006}
\APACinsertmetastar {%
Wilms2006}%
\begin{APACrefauthors}%
{Wilms}, J.%
, {Nowak}, M\BPBI A.%
, {Pottschmidt}, K.%
, {Pooley}, G\BPBI G.%
\BCBL {}\ \BBA {} {Fritz}, S.%
\end{APACrefauthors}%
\unskip\
\newblock
\APACrefYearMonthDay{2006}{{\APACmonth{02}}}{},
\newblock
\unskip
\newblock
\APACjournalVolNumPages{\aap}{447}{1}{245-261}.
\newblock
\begin{APACrefDOI} \doi{10.1051/0004-6361:20053938} \end{APACrefDOI}
\PrintBackRefs{\CurrentBib}

\bibitem [\protect \citeauthoryear {%
{Zakamska}%
\ \BBA {} {Greene}%
}{%
{Zakamska}%
\ \BBA {} {Greene}%
}{%
{\protect \APACyear {2014}}%
}]{%
Zakamska2014}
\APACinsertmetastar {%
Zakamska2014}%
\begin{APACrefauthors}%
{Zakamska}, N\BPBI L.%
\BCBT {}\ \BBA {} {Greene}, J\BPBI E.%
\end{APACrefauthors}%
\unskip\
\newblock
\APACrefYearMonthDay{2014}{{\APACmonth{07}}}{},
\newblock
\unskip
\newblock
\APACjournalVolNumPages{\mnras}{442}{1}{784-804}.
\newblock
\begin{APACrefDOI} \doi{10.1093/mnras/stu842} \end{APACrefDOI}
\PrintBackRefs{\CurrentBib}

\bibitem [\protect \citeauthoryear {%
{Zdziarski}%
, {Dzie{\l}ak}%
, {De Marco}%
, {Szanecki}%
\BCBL {}\ \BBA {} {Nied{\'z}wiecki}%
}{%
{Zdziarski}%
\ \protect \BOthers {.}}{%
{\protect \APACyear {2021}}%
}]{%
Zdziarski2021}
\APACinsertmetastar {%
Zdziarski2021}%
\begin{APACrefauthors}%
{Zdziarski}, A\BPBI A.%
, {Dzie{\l}ak}, M\BPBI A.%
, {De Marco}, B.%
, {Szanecki}, M.%
\BCBL {}\ \BBA {} {Nied{\'z}wiecki}, A.%
\end{APACrefauthors}%
\unskip\
\newblock
\APACrefYearMonthDay{2021}{{\APACmonth{03}}}{},
\newblock
\unskip
\newblock
\APACjournalVolNumPages{\apjl}{909}{1}{L9}.
\newblock
\begin{APACrefDOI} \doi{10.3847/2041-8213/abe7ef} \end{APACrefDOI}
\PrintBackRefs{\CurrentBib}

\bibitem [\protect \citeauthoryear {%
{Zdziarski}%
\ \protect \BOthers {.}}{%
{Zdziarski}%
\ \protect \BOthers {.}}{%
{\protect \APACyear {1998}}%
}]{%
Zdziarski1998}
\APACinsertmetastar {%
Zdziarski1998}%
\begin{APACrefauthors}%
{Zdziarski}, A\BPBI A.%
, {Poutanen}, J.%
, {Mikolajewska}, J.%
, {Gierlinski}, M.%
, {Ebisawa}, K.%
\BCBL {}\ \BBA {} {Johnson}, W\BPBI N.%
\end{APACrefauthors}%
\unskip\
\newblock
\APACrefYearMonthDay{1998}{{\APACmonth{12}}}{},
\newblock
\unskip
\newblock
\APACjournalVolNumPages{\mnras}{301}{2}{435-450}.
\newblock
\begin{APACrefDOI} \doi{10.1046/j.1365-8711.1998.02021.x} \end{APACrefDOI}
\PrintBackRefs{\CurrentBib}

\bibitem [\protect \citeauthoryear {%
{Zhu}%
\ \protect \BOthers {.}}{%
{Zhu}%
\ \protect \BOthers {.}}{%
{\protect \APACyear {2020}}%
}]{%
Zhu2020}
\APACinsertmetastar {%
Zhu2020}%
\begin{APACrefauthors}%
{Zhu}, S\BPBI F.%
, {Brandt}, W\BPBI N.%
, {Luo}, B.%
, {Wu}, J.%
, {Xue}, Y\BPBI Q.%
\BCBL {}\ \BBA {} {Yang}, G.%
\end{APACrefauthors}%
\unskip\
\newblock
\APACrefYearMonthDay{2020}{{\APACmonth{07}}}{},
\newblock
\unskip
\newblock
\APACjournalVolNumPages{MNRAS}{496}{1}{245-268}.
\newblock
\begin{APACrefDOI} \doi{10.1093/mnras/staa1411} \end{APACrefDOI}
\PrintBackRefs{\CurrentBib}

\end{thebibliography}

\end{document}